\documentclass[final]{l4dc2026}

\usepackage{microtype}
\usepackage{graphicx}

\usepackage{caption}
\usepackage{subcaption}

\usepackage{hyperref}

\usepackage[utf8]{inputenc} 
\usepackage[T1]{fontenc}    
\usepackage{hyperref}       
\usepackage{url}            
\usepackage{booktabs}       
\usepackage{amsfonts}       
\usepackage{nicefrac}       
\usepackage{microtype}      
\usepackage{xcolor}         

\usepackage{amsmath}
\usepackage{amssymb}
\usepackage{mathtools}

\usepackage{bbm}
\usepackage{enumitem}
\usepackage{siunitx}

\usepackage{algorithm}
\usepackage{algorithmic}

\usepackage{wrapfig}

\usepackage[capitalize,noabbrev]{cleveref}

\title[Select-DPC]{Choose Wisely: Data-driven Predictive Control for Nonlinear Systems Using Online Data Selection}
\usepackage{times}

\author{
  \Name{Joshua Näf} \Email{naefjo@ethz.ch} \\
  \Name{Keith Moffat} \Email{kmoffat@ethz.ch} \\
  \Name{Jaap Eising} \Email{jeising@ethz.ch} \\
  \Name{Florian Dörfler} \Email{dorfler@ethz.ch} \\
  \addr ETH Zurich, Zurich, Switzerland}

\begin{document}

\maketitle
\vspace{-0.4cm}

\begin{abstract}%
This paper proposes Select-Data-driven Predictive Control (Select-DPC), a new method for controlling nonlinear systems using output-feedback for which data are available but an explicit model is not. 
At each timestep, Select-DPC employs only the most relevant data to implicitly linearize the dynamics in ``trajectory space.'' Then, taking user-defined output constraints into account, it makes control decisions using a convex optimization. This data-driven optimal control is applied in a receding-horizon manner.
As the online data-selection is the core of Select-DPC, we propose and compare norm-based and manifold-embedding-based data selection methods. We evaluate Select-DPC on three benchmark nonlinear system simulators---rocket-landing, a robotic arm, and cart-pole inverted pendulum swing-up---comparing them with standard Data-enabled Predictive Control (DeePC) and Time-Windowed DeePC methods, and find that Select-DPC
outperforms both methods.
The source code can be found at: \url{https://github.com/naefjo/choose-wisely-paper}
\end{abstract}


\section{Introduction}
\begin{wrapfigure}[20]{r}{0.475\textwidth}
    \vspace{-1.6cm}
    \begin{center}
    \includegraphics[trim={0.2cm 0.2cm 0.2cm 0.2cm},clip,scale=1.05]{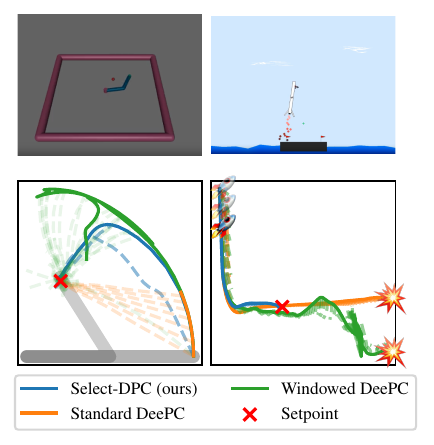}
    \vspace{-0.7cm}
    \caption{Closed-loop trajectories of Select-DPC (ours), standard DeePC, and Windowed DeePC in the Reacher \textit{(left)} and the Rocket Lander \textit{(right)} environments, as well as the corresponding open-loop predictions (dashed). Only Select-DPC converges to the setpoint.}
    \label{fig:title-figure}
    \end{center}
\end{wrapfigure} 
\vspace{-0.3cm}
The data-driven control and Reinforcement Learning (RL) communities are focused on controlling systems for which data is available, but a system model is not. 
Specifically, the control community has developed interest in ``direct'' Data-driven Predictive Control (DPC) methods \cite{dorfler_bridging_2023} as they reduce system modeling/identification time and effort and bring the benefits of predictive control to the ``model-free'' setting. 

This paper proposes Select-DPC, a direct DPC method for nonlinear systems.
Select-DPC extends DPC to the nonlinear setting by iteratively optimizing a convex Quadratic Program (QP) over an implicit linearization in ``trajectory space'' constructed from data.
We define trajectory space as the high-dimensional space in which each dimension corresponds to an input or output over the course of a discrete-time, finite-length trajectory.
Linearizing in trajectory space is achieved by selecting only trajectories that are ``close'' to the operating point in trajectory space.

\paragraph{Literature Review} 
Output-constrained systems are challenging for RL methods, which learn the constraints by exhaustive trial and error \cite{sutton_reinforcement_1998}.
Such systems have traditionally been the domain of Model Predictive Control (MPC) \cite{morari_model_1999}, which solves a receding-horizon optimal control problem based on a state-space representation with an online optimization.
In the nonlinear domain, the optimal control problem can be solved using Sequential Quadratic Programming (SQP). SQP-MPC iteratively linearizes the dynamics around a solution estimate, solves a QP around the linearization, and updates the solution estimate \cite{diehl2002real}.

\begin{wrapfigure}[16]{r}{0.4\textwidth}
    \centering
    \vspace{-0.4cm}
    \includegraphics[trim={1cm 0.9cm 0 0.9cm},clip,scale=0.8]{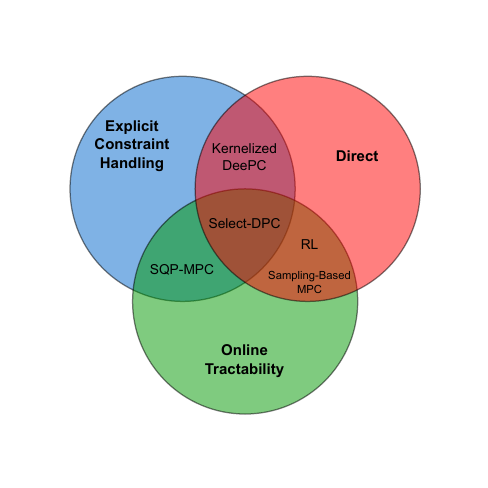}
    \vspace{-0.5cm}
    \caption{
    Venn diagram illustrating how the discussed methods fall within the \emph{Explicit Constraint Handling}, \emph{Direct}, and \emph{Online Tractability} merit categories. Select-DPC has all three merits.
    }
    \label{fig:venn}
\end{wrapfigure}
The class of Data-driven Predictive Controllers construct feasible trajectories of a linear, time-invariant (LTI) system purely from data. 
These methods bypass the sequential \emph{indirect} system identification and MPC design pipeline and are hence called \emph{direct methods}.
Data-enabled Predictive Control (DeePC) \cite{coulson_data-enabled_2019}, for instance, linearly combines input-output trajectories in the data set. 
Such predictions fit into the behavioral control framework \cite{willems_paradigms_1991}. For a survey on DeePC we refer the reader to \cite{markovsky_data-driven_2023}.
DPC methods have demonstrated impressive results on various control tasks. While DPC methods can handle minor violations of the LTI assumption by treating them as additive measurement noise, the LTI restriction is severe and leads to arbitrary performance loss on generic nonlinear systems. 

Several approaches to DeePC in the nonlinear case can be found in the literature.
\citet{berberich_overview_2024} provide tailored formulations of the Fundamental Lemma \cite{willems_note_2005} to specific classes of nonlinear systems.
These formulations enable efficient synthesis of DPC for systems belonging to these specific
system classes. However, this already imposes a strong prior on the global nonlinear structure of the dynamical process,
which might not be available for an arbitrary system.
Alternatively, \citet{berberich_linear_2022}
propose using a set of the most recent past observations in a sliding window fashion along the current closed-loop trajectory to form the basis of a linear predictor, herein referred to as Time-Windowed DeePC.
As this method only uses recent closed-loop data,
it is practical only for systems with benign nonlinearities and with high signal-to-noise ratios.
Approaches in the literature also suggest to 
reformulate the predictor in a kernelized fashion, enabling
the use of nonlinear kernels \cite{pmlr-v144-lian21a, huang_robust_2024}, using nonlinear basis functions or reformulating the problem in a lifted state-space using Koopman operators \cite{lazar_basis_2023}.
While these methods offer nonlinear function approximation, the resulting
optimization becomes high-dimensional, nonlinear, and generally nonconvex, requiring increased online computational complexity and resulting in suboptimal outcomes.

In contrast with the optimization-based MPC literature, sampling-based MPC methods generate an ensemble of roll-outs by randomly sampling control inputs \cite{nagabandi_neural_2018} and then applying an iterative refinement to obtain near optimal control sequences \cite{botev_chapter_2013, williams_model_2017}.
These methods are favored in domains such as RL, as sampling-based MPC only requires zero-order model information, i.e. no gradient of the model, making them computationally cheap to evaluate.
However, sampling-based MPC usually has no structured way of incorporating output constraints, since they rely on model roll-outs, and require the user to offload this complexity into other parts of the method, such as the model itself or the cost.
In the context of RL, optimization-based planning techniques with explicit, learning-based models suffer from accumulating prediction errors \cite{moerland_model-based_2022}, exploding gradients due to backpropagation through time or model exploitation in regions of low data density \cite{kurutach_model-ensemble_2018}.
\Cref{fig:venn} shows a conceptual overview of the mentioned strategies and their respective traits. 

The Select-DPC data-selection methods rely on finding high-dimensional nearest-neighbors, a concept that has been well established and applied in adjacent fields such as SLAM for feature matching and pose optimization \cite{bailey2006simultaneous}, 3D shape registration \cite{besl1992method} and in the context of dimensionality reduction and manifold learning \cite{jia2022feature}.


\paragraph{Contributions}
\vspace{-0.2cm}
The contributions of this paper are threefold:
\vspace{-0.2cm}
\begin{enumerate}[leftmargin=0.6cm]
    \item We introduce \emph{Select-DPC}, a data-driven quadratic program predictive controller for nonlinear systems which implicitly linearizes the system dynamics in trajectory space by selecting and using only the most relevant data at each decision moment.
    \vspace{-0.3cm}
    \item We propose two data-selection methods to achieve linearization in trajectory space, a norm-based method and a manifold-embedding-based method.
    \vspace{-0.3cm}
    \item We demonstrate the benefits of Select-DPC on three benchmark nonlinear systems including a planar rocket-landing, a two degree of
    freedom robotic arm, and a cart-pole inverted pendulum, comparing it to standard DeePC and a Time-Windowed DeePC approach
    and find that Select-DPC outperforms both in terms of closed-loop cost and prediction accuracy.
\end{enumerate}

\vspace{-0.6cm}
\section{Background}
\vspace{-0.1cm}
\paragraph{Preliminaries} We concern ourselves with nonlinear systems, for which we assume an autoregressive form $y_f(k) = f(y_\text{p}(k), u_\text{p}(k), u_\text{f}(k))$
exists.
We use $y\in\mathbb{R}^p$ and $u\in\mathbb{R}^m$ to denote measurements and inputs of the nonlinear system respectively.
The subscripts $a_\text{p}$ and $a_\text{f}$ denote past and future values of a variable $a$ and are defined as $a_\text{p} (k) \coloneqq \{a(k-i)\}_{i=T_\text{p}-1}^0$ and $a_\text{f} (k) \coloneqq \{a(k+i)\}_{i=1}^{T_\text{f}}$.

\vspace{-0.1cm}
\paragraph{Data-Enabled Predictive Control (DeePC)}
Select-DPC builds upon linear DPC which solves a receding-horizon predictive control problem subject to input and output constraints. The novelty of DPC is that it achieves predictive control purely based on implicit predictions of input-output data without ever constructing an explicit predictor (a model) of future measurements given a set of inputs. 
DeePC \cite{coulson_data-enabled_2019}, for instance, solves the following optimization problem
\vspace{-0.2cm}
\begin{equation}
    \min_{y,u,g} \sum_{i=1}^{T_f} c_i(y_i, u_i) + r(g),\ \text{s.t.}\ Hg = \left[u_\text{p}^\top, y_\text{p}^\top, u^\top, y^\top\right]^\top,\ (u_i,y_i) \in \mathcal{U}\times\mathcal{Y}\ \forall i = \{1, \dots T_f\},
\vspace{-0.2cm}
\end{equation}
where each column in the matrix $H$ represents a $L$-length trajectory from a pre-collected data set $\mathcal{D} = \{\tau_0, \dots, \tau_{n_d}\}$, with $\tau = \{(u_i, y_i)\}_{i=1}^L$. $H$ can be partitioned into $\left[ H_\text{p}^\top, H_\text{f}^\top\right]^\top$, where each factor represents the partition of the data set into past and future quantities respectively. 
Several regularizer $r(g)$ have been introduced to further bias the choice of said linear combinations \cite{dorfler_bridging_2023}. We refer to \cite{markovsky_data-driven_2023} or \cref{app:dpc} for a detailed discussion of DeePC.

Solving DeePC online using a large data set of trajectories is intractable as the solve time of the
optimization problem scales superlinearly with $\operatorname{card}(\mathcal{D})$. Furthermore, in the presence of
nonlinearities, the LTI assumption of DeePC is violated, resulting in suboptimal performance.

\vspace{-0.4cm}
\section{Select-DPC}\label{sec:select-dpc}
\vspace{-0.2cm}
We now introduce \emph{Select-DPC}, an algorithm that allows us to solve a nonlinear data-driven receding-horizon
optimal control problem using a convex quadratic programming routine by choosing, at each solver iteration, a subset of trajectories $\tilde{\mathcal{D}}$ with cardinality
$\operatorname{card}(\tilde{\mathcal{D}}) = N_\text{cols}$ from $\mathcal{D}$ where ${N_\text{cols}\ll \operatorname{card}\left(\mathcal{D}\right)}$.
\Cref{alg:select-dpc} summarizes the implementation of Select-DPC.
In between two
\begin{wrapfigure}[10]{r}{0.5\textwidth}
\vspace{-0.7cm}
\begin{minipage}{0.5\textwidth}
\begin{algorithm}[H]
\caption{Select-DPC}\label{alg:select-dpc}
\begin{algorithmic}[1]
  \FUNCTION{Select-DPC($u_\text{p}$, $y_\text{p}$)}
    \WHILE{convergence criterion not met}
      \STATE $\tilde{\tau} \gets \texttt{DPC.getLastPrediction}()$
      \STATE $\tilde{D} \gets \texttt{select}(\mathcal{D},\, \tilde{\tau},\, N_\text{cols})$ \label{alg:select-dpc:select}
      \STATE $u \gets \texttt{DPC.genAction}(u_\text{p},\ y_\text{p},\ \tilde{D})$\label{alg:select-dpc:solve}
    \ENDWHILE
    \STATE Return $u_0$
  \ENDFUNCTION
\end{algorithmic}
\end{algorithm}
\end{minipage}
\end{wrapfigure}
real-time sampling instances, an iterative procedure is performed where, at every iteration, we select from a data set $\mathcal{D}$ a subset of trajectories $\tilde{\mathcal{D}}$ that are ``relevant'' to the current linearization point given by the open-loop solution $\tilde{\tau}$. While we loosely use the term relevance here, this mechanism will be discussed in more detail in \cref{sec:data-sel}.
Select-DPC then solves a DPC problem using the subset of data. This is iterated until some convergence criterion, e.g. specified tolerance or maximum number of iterations, is reached. 
At the first time step, the open-loop trajectory is warm-started by uniformly replicating the initial system measurement across the horizon, with zero inputs assumed.
Notice that there is no restriction on the choice of DPC solver.
We now discuss a number of important attributes of this proposed method.

\vspace{-0.2cm}
\paragraph{Cost-agnostic} Since the data set generation is structurally independent of the control formulation, a posteriori cost or constraint modification is possible, allowing the method to be applied zero-shot to new problems without the need for additional data collection.

\vspace{-0.2cm}
\paragraph{Hyperparameters} The performance of {Select-DPC} depends on the tuning of several hyperparameters. First, the DPC specific hyperparameters, such as regularization terms and horizon lengths $T_\text{p},\ T_\text{f}$. These can either be hand-tuned according to established heuristics \cite{elokda_data-enabled_2021} or optimized through gradient-descent methods \cite{cummins_deepc-hunt_2024} or black-box parameter tuning \cite{berkenkamp_safe_2016}. For DPC tuning guidelines, we refer to \cite{markovsky_data-driven_2023}. Select-DPC adds onto these the design parameters $N_\text{cols}$ as well as the method of data selection. In the following section, we will investigate two different data-selection methods which subsample a specified number of trajectories from a large offline library of trajectories observed during the data gathering process.

\vspace{-0.2cm}
\paragraph{Sequential Linearization} The iterative refinement of the solution until convergence is comparable to the model-based SQP-MPC algorithm for nonlinear systems \cite{diehl2002real}.
We note that Select-DPC's implicit linearization is performed in trajectory space. In the model-based case, linearizing in trajectory space corresponds to a linear time-varying parametrization of the nonlinear dynamics for the MPC. This linear, time-varying parametrization  contrasts with an LTI formulation in which the Jacobians are only evaluated at the current measurement $y(k)$ and kept fixed along the horizon. See \cref{app:sqpmpc} for more details.

\vspace{-0.3cm}
\section{Data Selection}\label{sec:data-sel}
\vspace{-0.3cm}
At the heart of Select-DPC lies the data selection method, which selects data points (representing trajectories $\tau_i$) from a large (offline) data set. Since the subproblem solved by Select-DPC relies on linearly combining trajectories, the subset of data should exhibit predominantly linear dynamics.
The suggested approach in the literature is to use a sliding window over past input-output measurements. We propose that $\tilde{\mathcal{D}}$ should be chosen according to some \emph{spatial distance metric} (in trajectory space) as opposed to \emph{temporal proximity} (such as a sliding window) to the current linearization point $\tilde{\tau}$.

In the following, we will present norm-based and manifold-embedding-based selection, while placing emphasis on the modular nature of Select-DPC. While norm-based selection is efficient to implement and allows the user to easily update the data set online, manifold-embedding-based selection enables the distance computation in a lower dimensional embedding space which lessens the impact of the \emph{curse of dimensionality} but comes with increased offline computation cost.

\vspace{-0.2cm}
\paragraph{Norm-based selection}
\begin{wrapfigure}[6]{r}{0.47\textwidth}
\vspace{-0.8cm}
\begin{minipage}{0.47\textwidth}
\begin{algorithm}[H]
  \caption{Norm-based Data Selection}\label{alg:norm-selection}
  \begin{algorithmic}[1]
    \FUNCTION{NormDataSelection($\mathcal{D}$, $\tilde{\tau}$, $N_\text{cols}$)}
      \STATE $\tilde{\mathcal{D}}$ =  \texttt{sort}$\left(\{\|\tau_i - \tilde{\tau}\|\, |\, i = 1,\dots,n_d\}\right)$ 
      \STATE Pick first $[1, N_\text{cols}]$ from $\tilde{\mathcal{D}}$
    \ENDFUNCTION
  \end{algorithmic}
\end{algorithm}
\end{minipage}
\end{wrapfigure}
Norm-based data selection selects from a data set $\mathcal{D}$ the $N_\text{cols}$ closest (in space) trajectories by computing the norm of the differences between the open-loop solution $\tilde{\tau}$ and the trajectories in the data set $\mathcal{D}$. An implementation is given in \cref{alg:norm-selection}.

While this approach is intuitive, it is common knowledge that norm-based distances fall flat in high dimensional spaces due to the \emph{curse of dimensionality} \cite{goos_surprising_2001}. Norm-based selection computes distances of trajectories, i.e. vectors with dimension ${(T_\text{p}+T_\text{f})(m+p)}$ and is hence also susceptible to this curse.

\vspace{-0.2cm}
\paragraph{Low Dimensional Representation using Manifold Learning}
\begin{wrapfigure}[8]{r}{0.5\textwidth}
\vspace{-0.8cm}
\begin{minipage}{0.5\textwidth}
\begin{algorithm}[H]
  \caption{Manifold-Embedding Data Selection}\label{alg:isomap-select}
  \begin{algorithmic}[1]
    \FUNCTION{ManifoldDataSel($\mathcal{D}$, $\tilde{\tau}$, $N_\text{cols}$)}
    \STATE $\mathcal{D}_\text{e} \gets \texttt{embed}\left(\mathcal{D}\right)$\label{alg:isomap-select:dataset}
    \STATE $\tilde{\tau}_\text{e} \gets \texttt{embed}\left(\tilde{\tau}\right)$
      \STATE $\tilde{\mathcal{D}_\text{e}}$ =  \texttt{sort}$\left(\{\|\tau_{\text{e},i} - \tilde{\tau}_\text{e}\|\, |\, i = 1,\dots,n_d\}\right)$ 
      \STATE Pick first $[1, N_\text{cols}]$ from $\tilde{\mathcal{D}}_\text{e}$
    \ENDFUNCTION
  \end{algorithmic}
\end{algorithm}
\end{minipage}
\end{wrapfigure}
The field of manifold learning (nonlinear dimensionality reduction) concerns itself with the compression of high-dimensional data onto manifolds with lower intrinsic dimension than the original data by using (nonlinear) projection techniques \cite{lee_nonlinear_2007, jia2022feature}.
Manifold Learning techniques rely on the \emph{manifold hypothesis} \cite{fefferman_testing_2016}, which postulates that the intrinsic dimensionality of a data set $\mathcal{D}$ is much lower than the dimensionality of the individual data points.
For instance, if the system under consideration is a controllable and deterministic LTI  system, then Willems' Fundamental Lemma states
that the trajectories span a manifold (in particular a subspace) that can at most have $n + T_\text{f}\cdot m$ degrees of freedom \cite{willems_note_2005}, which is generally less than the trajectory
dimensionality $(T_\text{p} + T_\text{f})(m+p)$. Here, $n$ denotes the dimension of the process's latent state, $p$ is the dimension of the measurement vector and $m$ indicates the dimension of the input signal. Motivated by this observation, we wish to find a lower dimensional
representation of the data where the similarities can be computed without incurring the curse of dimensionality.  

Isomap, introduced by \citet{tenenbaum_global_2000}, is an unsupervised manifold learning method which aims to find an
embedding of the data in a Euclidean space which preserves geodesic distances of a neighborhood graph of the data
constructed in the data space.
This comes with increased computational cost compared to norm based data selection but can offer inter data point distances which more accurately reflect the nonlinear nature of the data manifold in trajectory space. Also notice that the bulk of this added cost is a one-time upfront cost (computing the embedding), which can be performed offline.

Although this does not fully resolve the problem caused by the curse of dimensionality as the embedding dimension,
i.e. the dimension where we need to compute distances, will still scale with at least with
$n + T_\text{f}\cdot m$, it does offer a structured way of compressing the data to a reasonable dimension.
Experiments showed that embedding dimensions lower than the minimum dimension required to represent a linear system already
lead to diminishing returns in reconstruction error.

The pseudocode for manifold-embedding-based data selection using the Isomap algorithm is outlined in \cref{alg:isomap-select}. For algorithmic details of Isomap (i.e. the \texttt{embed} step) we refer to \cite{tenenbaum_global_2000} or \cref{app:isomap}.
Notice that the embedding of the data set in line 2 can be
cached/computed offline. Online, the algorithm then proceeds to embed the query trajectory $\tilde{\tau}_i$ and computes its similarities \emph{in the lower dimensional embedding space}. Then, similarly to \cref{alg:norm-selection}, the relevant trajectories are returned in the trajectory space, after which the implicit predictor for DPC is constructed using the retrieved non-embedded data points.

\vspace{-0.5cm}
\section{Results and Discussion}\label{sec:results}
\begin{figure}[t]
  \centering
    \begin{minipage}{0.45\textwidth}
        \centering
        \includegraphics[width=0.9\linewidth]{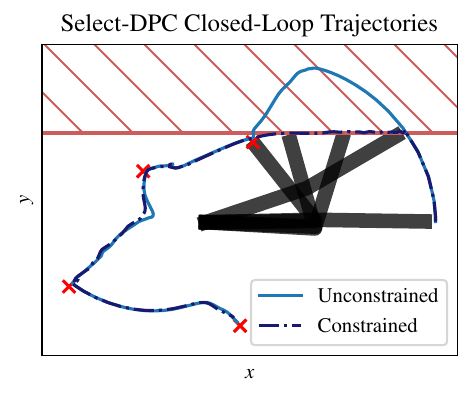}
    \end{minipage}\hfill
    \begin{minipage}{0.45\textwidth}
        \centering
        \includegraphics[width=\linewidth]{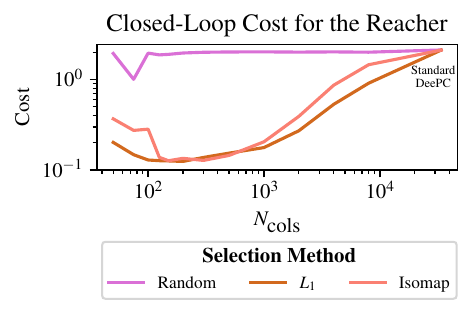}
    \end{minipage}
    \vspace{-0.6cm}
    \caption{\emph{(left)} Closed-loop trajectories for a constrained (dark blue) and unconstrained (blue) DPC subproblem.
    \emph{(right)} Comparing the closed-loop cost incurred as a function of subset cardinality. Clearly there is a sweet spot in how many trajectories should be selected for the predictions. 
    }
    \label{fig:reacher_figures}
    \vspace{-0.6cm}
\end{figure}

\vspace{-0.3cm}
Simulation experiments were performed to validate the efficacy of Select-DPC and to benchmark it
against standard DeePC \cite{coulson_data-enabled_2019}, which utilizes the entire data set, as well as the Time-Windowed DeePC variant proposed by \citet{berberich_linear_2022},
which constructs the implicit predictor by continuously updating the data set with new input-output data and discarding the most out-of-date measurement. All experiments were performed on an Apple MacBook Pro with an M2 Pro chip and 16GB of RAM. We use a maximum number of iterations $n_\text{max}$ as a stopping criterion.
No process or measurement noise is added to the simulated systems.
Hyperparameters are provided in \cref{app:hyperparams}. In the experiments focusing on closed-loop cost (\cref{fig:rows_vs_costs_solve_times} left and \cref{fig:reacher_figures} right), we performed an LQ decomposition of the data matrix \cite{breschi_data-driven_2023} for data subsets larger than 400. While this drastically speeds up computation times, we did not observe significant differences in control performance compared to no LQ decomposition.

\vspace{-0.3cm}
\subsection{Landing a Reusable Rocket}
We use a planar vertical takeoff, vertical landing rocket to demonstrate the performance of Select-DPC on an open-loop unstable nonlinear system. We show that Select-DPC successfully copes with the nonlinear dynamics and manages to stabilize the system at a given setpoint.
A custom environment for the gymnasium suite
\cite{kwiatkowski_gymnasium_2024} was used as a simulation platform.
For a detailed description of the environment, we refer to \cite{cummins_deepc-hunt_2024}.

\begin{wrapfigure}[14]{r}{0.45\textwidth}
  \vspace{-0.75cm}
  \begin{center}
    \includegraphics[trim={0cm 0.15cm 0 0.15cm},clip,scale=0.85]{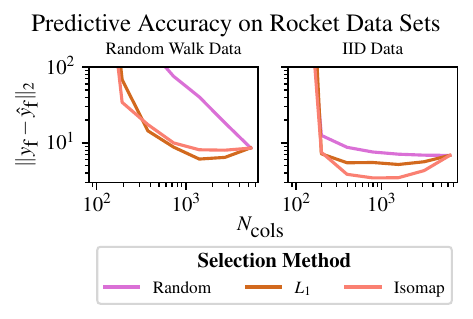}
  \end{center}
  \vspace{-0.8cm}
  \caption{Residual between the ground-truth and least-squares predictions as a function of $N_\text{cols}$ used in the predictor.
  Rightmost data point corresponds to standard DeePC.}
  \label{fig:prediction_validation}
\end{wrapfigure}
The right column in \cref{fig:title-figure} shows the simulator setup and specifically, in the bottom figure, three closed-loop rocket trajectories.
One for Select-DPC, one for standard DeePC using the full data set and one for Time-Windowed DeePC respectively.
We use the Isomap-based selection method and ${N_\text{cols}=500}$.
While both the standard DeePC and the Time-Windowed DeePC controllers diverge,
Select-DPC manages to regulate the system to the desired setpoint and stabilize it there.

Furthermore, we compare the performance of Select-DPC using the two presented selection methods on two data sets.
Both data sets were collected using input sequences of the form $u_{t} = a\cdot u_{t-1} + b\cdot n,$ with $ n \sim \mathcal{U}(-1, 1)$.
The data set \emph{IID Data} uses $(a, b) = (0, 1)$ and \emph{Random Walk Data} uses $(a, b) = (1, 0.1)$. 
To ensure that a large region of
the trajectory space was covered by the collected data, 100 simulations were run with randomized initial conditions and a maximum of 100 timesteps each, resulting in
data sets with  approximately $\num{5000}$ data points.

\vspace{-0.3cm}
\paragraph{Ablation Study}
The comparisons of Select-DPC on the two data sets are two-fold. First, the resulting predictive accuracy of the selection methods are compared. Furthermore, a comparison of the realized closed-loop cost is given as a function of selection method and data set. As a benchmark, random sketching of the full data set at each decision moment is used.

The predictive accuracy of the methods was compared by computing the least squares solution of the predictor used in the DPC subproblem as a function of the number $N_\text{cols}$ of trajectories in the selected set of data. 
For each trajectory in the holdout data set, constructed from trajectory-segments of a closed-loop trajectory, the $N_\text{cols}$ closest trajectories (in spatial $L_1$-norm or using Isomap-embedding) under the given selection method were
selected and the prediction of future measurements $\hat{y}_\text{f}$, given the corresponding future input sequence $u_\text{f}$ were computed using the least square predictor of $Hg = [u; y]$. 
The plots in \cref{fig:prediction_validation} show the cumulative residuals between the predicted future trajectory
and the ground truth realization $\|\hat{y}_\text{f} - y_\text{f}\|_2$ over the entire validation set.
\begin{figure}[t]
    \centering
    \begin{minipage}{0.45\textwidth}
        \centering
        \includegraphics[width=\linewidth]{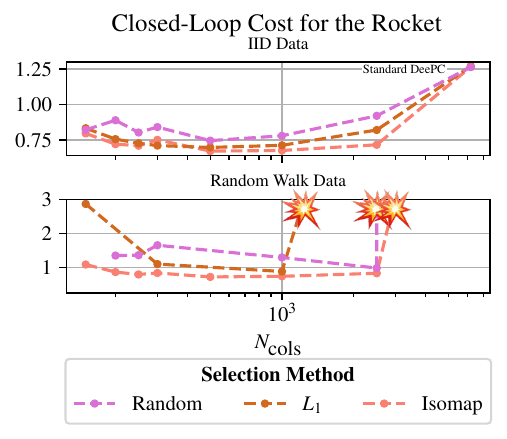}
    \end{minipage}\hfill
    \begin{minipage}{0.45\textwidth}
        \centering
        \includegraphics[width=\linewidth]{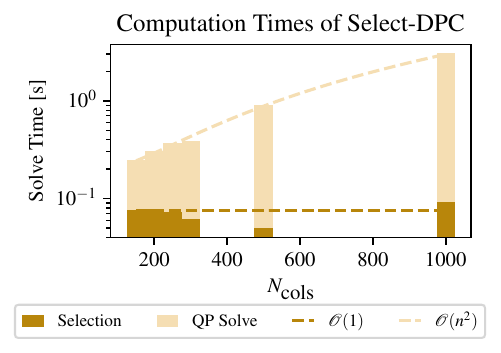}
    \end{minipage}
    \vspace{-0.4cm}
    \caption{\emph{(left)} Closed-loop cost as a function of data subset cardinality on two different data sets. Instances where the selection methods destabilized the system (i.e. infinite cost) have been omitted and the remaining finite cost data points are shown.
    \emph{(right)} Solve times of Select-DPC as a function of $N_\text{cols}$ decomposed into data selection and QP solve time. DeePC is used to solve the QP subproblem.
    }
    \label{fig:rows_vs_costs_solve_times}
    \vspace{-0.6cm}
\end{figure}
In the extreme case where $N_\text{cols}=\operatorname{card}(\mathcal{D})$, each of the methods coincides with the least squares solution of the predictor in standard DeePC. We observe that both methods of data selection (based on spatial $L_1$-norm or Isomap-embedding) lead to lower cumulative residual on the holdout validation set when compared to using the full data set or random sketching. In particular, there is a sweet spot in the number $N_\text{cols}$ of trajectories to use in the predictor.

The left plot of \cref{fig:rows_vs_costs_solve_times} shows the closed-loop performance of Select-DPC on the setpoint tracking task depicted in the right column of \cref{fig:title-figure} on both IID Data and Random Walk Data.
We again observe on both data sets that selecting data outperforms using the full data set (i.e. standard DeePC) in terms of closed-loop performance,
as a lower cost could be achieved.
Using IID Data, the reduction in closed-loop
cost is 1.9. 
Standard DeePC, using the full Random Walk data set, failed to stabilize the rocket.

The level of nonlinearity in the data has a significant effect on the performance of standard DeePC.
Due to the high inertia of the rocket, the IID Data predominantly exhibits behaviors that can be represented reasonably well by a linear model, and hence, standard DeePC can be applied. However, this comes at a performance penalty as nonlinear effects are not accounted for. We observe that Select-DPC outperforms standard DeePC on IID Data and that data selection can still lead to performance increases if the data set does not fully capture the nonlinear behavior of the underlying system.
The random walk structure of the input sequence in Random Walk Data, on the other hand,
ensures that the nonlinearities of the system are sufficiently excited. This makes standard DeePC fail but provides Select-DPC with more global information about the nonlinear system.

\vspace{-0.3cm}
\paragraph{Computational Cost}
The right plot of \cref{fig:rows_vs_costs_solve_times} shows the computation times of Select-DPC (without LQ decomposition) as a function of number of subselected trajectories. The dominant factor in the solve times is solving the quadratic program, which empirically scales at least quadratically with $N_\text{cols}$, while data selection is a constant time operation with respect to $N_\text{cols}$. 
Concluding from \cref{fig:rows_vs_costs_solve_times}, reducing the number of trajectories is beneficial from both a computational and a closed-loop performance standpoint.
Select-DPC not only increases performance in the nonlinear domain but also reduces computational cost compared to standard DeePC by one order of magnitude.

\vspace{-0.8cm}
\subsection{Planar Robotic Manipulator}
\vspace{-0.2cm}
We use the \emph{Reacher} environment \cite{kwiatkowski_gymnasium_2024}, which simulates a planar robotic manipulator, to demonstrate the ability of Select-DPC to control a highly nonlinear system and successfully track reference setpoints. Furthermore, we show that Select-DPC is effective in coping with and respecting a posteriori specified output constraints.
The kinematic chain of the manipulator consists of two rigid bodies that are connected by revolute joints which are
each actuated by a torque, giving it an action space of $u \in \mathbb{R}^2$. The measurement vector consists of the
sine and cosine of both joint angles respectively, and the end-effector position in $\mathbb{R}^2$.
Furthermore, we are also able to measure the angular velocity of each link, resulting in an observation space
$y \in \mathbb{R}^8$. This system is more challenging to control using DeePC (or any linear control method) since
the end-effector position is dictated by the composition of two rotations.
In this environment, it is not unreasonable to assume that the angles must exceed $90^\circ$
in order to reach a target. Furthermore, for a given end-effector position, the corresponding joint configuration is not
necessarily unique.

Select-DPC was successfully able to track an end-effector reference as shown in the left column of \cref{fig:title-figure}. We use $L_1$-based data selection and $N_\text{cols}=100$.
While Select-DPC is able to converge to the desired setpoint, standard DeePC fails to track the
reference during the entire simulation horizon. Clearly, the open-loop predictions do not align with the closed-loop
behavior, indicating that using the full data set results in bad predictive accuracy.
Time-Windowed DeePC also fails to converge to the setpoint.
We again investigate the effect of number of selected trajectories $N_\text{cols}$, as well as selection strategy on the resulting closed loop cost.
For the Reacher, data were collected using IID inputs. In total, 200 simulations were run with 200 steps each.
Here, only IID inputs were collected instead of random walk because
the system has lower inertia compared to the rocket and the nonlinearities could be excited more easily.

\vspace{-0.3cm}
\paragraph{Ablation Study}
The right plot of \cref{fig:reacher_figures} shows the incurred closed-loop cost as a function of number of trajectories used in the
implicit predictor. Similarly to the rocket, we observe a decrease in predictive accuracy and an increase
in cost as the number of data points used in the predictor increases. While random sketching showed decent results in the rocket simulation, it fails completely in the reacher simulation as no accurate predictions could be generated,
resulting in the controller not making any progress towards the setpoint and hence accumulating high closed-loop cost.

\vspace{-0.2cm}
\paragraph{Zero-Shot Generalizability of Select-DPC}
The left plot of \cref{fig:reacher_figures} shows two closed-loop trajectories of Select-DPC where the controller was tasked with tracking a set of consecutive end-effector references. 
One simulation was left unconstrained, while a constraint on $y$ was added to the DPC subproblem of the other one, indicated by the red region.
Since the DeePC formulation used in Select-DPC admits arbitrary, a priori unknown output-constraints,
we observe that Select-DPC is able to generalize to different tasks across the entire trajectory manifold of the Reacher by reusing the same data set and without structural changes to the controller.
The structure of DeePC effectively decouples the dynamics learning/training process from the controller design and Select-DPC extends this to the nonlinear case. Indeed, both the constrained and unconstrained tasks use the same data set which is unaware of the structure of the predictive control problem. Thanks to the DeePC subproblem, this allows us to easily adapt the cost function or constraint terms for the task at hand in zero-shot fashion without requiring task specific data or additional online learning.

\vspace{-0.3cm}
\subsection{Cart-Pole inverted pendulum swing-up}
\vspace{-0.2cm}
We show that Select-DPC is capable of performing a cart-pole inverted pendulum swing-up.
The input $u\in\mathbb{R}$ is a force acting on the cart along the $x$-axis. Furthermore, the measurement $y\in\mathbb{R}^5$ consists of the cart position, the sine and cosine of the angle between the pendulum and the vertical axis, as well as linear and angular velocities of the cart and pendulum respectively.

Due to the unstable nature of the setpoint and the system starting in a stable equilibrium position, data was gathered in closed-loop.
Specifically, the data set consists of 200 demonstrations of successful but suboptimal swing-ups, collected using a stabilizing controller \cite{CHATTERJEE2002355}.

\begin{figure}[t]
  \centering
  \begin{minipage}{0.45\textwidth}
        \centering
        \includegraphics[width=\linewidth]{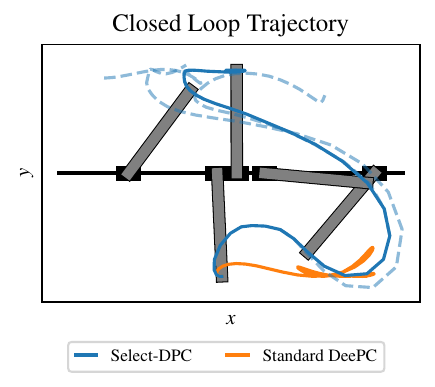}
        \label{fig:rows}
    \end{minipage}\hfill
    \begin{minipage}{0.45\textwidth}
        \centering
        \includegraphics[width=\linewidth]{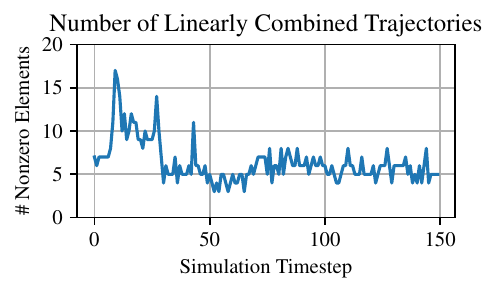}
        \label{fig:times}
    \end{minipage}
  \vspace{-0.9cm}
  \caption{
    \emph{(left)} Closed-loop trajectory for an inverted pendulum swing-up. Select-DPC (blue) is successful in transitioning from the downward facing stable equilibrium to the upward facing unstable equilibrium. Standard DeePC (orange) fails to swing up. The dashed lines indicate the open-loop prediction of Select-DPC at the four snapshot time instances, $t$=15, 22, 32 and 120. 
    \emph{(right)} Number of trajectories used to construct the final prediction at each sampling instant of Select-DPC.
  }
  \label{fig:inv_pend_figs}
  \vspace{-0.6cm}
\end{figure}

The left plot of \cref{fig:inv_pend_figs} shows the closed-loop trajectories of both Select-DPC and standard DeePC when tasked with tracking a setpoint at the upward facing, unstable equilibrium while starting from the downward facing stable equilibrium. While standard DeePC is unable to perform the swing-up and oscillates around the stable equilibrium, Select-DPC is successfully able to learn from the demonstrations in the data set and performs a swing-up to the desired vertical position and subsequently stabilizes the pendulum there. It is worthwhile to point out that the controller does not simply select the top trajectory in the selected data set. Instead, Select-DPC linearly combines several trajectory segments at each iteration of the solver to produce the swing-up (c.f. right plot of \cref{fig:inv_pend_figs}). This is also evident from the fact that the horizon in the predictive controller is shorter than the episode length, thus the solver never has access to a full swing-up trajectory.

\vspace{-0.5cm}
\section{Limitations}
\vspace{-0.3cm}
Controllers relying on online optimization inherently require increased computational time compared to other methods. The current computational cost of Select-DPC of 200-400 ms prevents it from being employed in highly dynamic applications.
A further bottleneck is Select-DPC's memory footprint. Large prediction horizons and large data sets could  violate memory constraints.
A limiting factor of Select-DPC, hindering a study of its performance on a wide range of control tasks, is the necessity of hyper parameter tuning to achieve satisfactory control performance. While techniques exist for automatic controller tuning in this context, they have not been applied in this study and a large-scale study of Select-DPC to empirically quantify its capabilities is subject to future work.

\vspace{-0.5cm}
\section{Conclusion}
\vspace{-0.3cm}
This paper introduces Select-DPC, a novel approach to constraint-aware Data-driven Predictive Control for nonlinear systems. At each time instance, the method pre-processes the data set to determine the most relevant data, then passes the most relevant data to a convex optimization which determines the optimal control for the future horizon. 
The optimal control for the given timestep is implemented, and the algorithm progresses in a receding horizon fashion.

In contrast to other DPC methods, Select-DPC can be implemented easily, scales favorably with the number of collected data-points and increases performance outside of the linear domain. Furthermore, compared to sampling-based MPC, Select-DPC can explicitly handle output constraints.

We validated this on a set of three simulation environments.
In the rocket simulator environment, we showed that for one of the data sets considered, Select-DPC cut closed-loop cost by a factor of 2.
For the other data set, our method resulted in stable closed-loop behavior, where previous methods crashed.
In the second set of simulations in the Reacher environment, we enabled zero-shot constrained setpoint tracking of a reference signal in a robotic reacher simulation.
A task unachievable by previous DPC formulations.
Finally, in the last set of simulations of a cart-pole inverted pendulum, we demonstrated successful transitioning between two different equilibria of the system by performing a pendulum swing-up.

A direction of future work is to investigate different selection methods, such as structure-exploiting selection methods or selection methods based on an information criterion. 
Another direction is to replace Select-DPC's large training data set with a data-driven forward simulator of the dynamics (i.e., a ``world model''), resulting in a constraint-aware alternative to sampling-based MPC.

\acks{
The authors would like to thank Xian Li for providing the simulation environment and data collection procedure for the inverted pendulum.
}

\bibliography{references}

@ARTICLE{besl1992method,
  author={Besl, P.J. and McKay, Neil D.},
  journal={IEEE Transactions on Pattern Analysis and Machine Intelligence}, 
  title={A method for registration of 3-D shapes}, 
  year={1992},
  volume={14},
  number={2},
  pages={239-256},
  doi={10.1109/34.121791}
}

@article{bailey2006simultaneous,
  author={Bailey, T. and Durrant-Whyte, H.},
  journal={IEEE Robotics \& Automation Magazine}, 
  title={Simultaneous localization and mapping (SLAM): part II}, 
  year={2006},
  volume={13},
  number={3},
  pages={108-117},
  doi={10.1109/MRA.2006.1678144}
}

@article{diehl2002real,
  title={Real-time optimization and nonlinear model predictive control of processes governed by differential-algebraic equations},
  author={Diehl, M. and Bock, H. G. and Schl{\"o}der, J. P and Findeisen, R. and Nagy, Z. and Allg{\"o}wer, F.},
  journal={Journal of Process Control},
  volume={12},
  number={4},
  pages={577--585},
  year={2002},
  publisher={Elsevier}
}

@article{CHATTERJEE2002355,
title = {Swing-up and stabilization of a cart–pendulum system under restricted cart track length},
journal = {Systems \& Control Letters},
volume = {47},
number = {4},
pages = {355-364},
year = {2002},
author = {D. Chatterjee and A. Patra and H. K. Joglekar}
}

@article{jia2022feature,
  title={Feature dimensionality reduction: a review},
  author={Jia, W. and Sun, M. and Lian, J. and Hou, S.},
  journal={Complex \& Intelligent Systems},
  volume={8},
  number={3},
  pages={2663--2693},
  year={2022},
  publisher={Springer}
}

@book{lee_nonlinear_2007,
	address = {New York, NY},
	series = {Information {Science} and {Statistics}},
	title = {Nonlinear {Dimensionality} {Reduction}},
	copyright = {http://www.springer.com/tdm},
	isbn = {978-0-387-39350-6},
	language = {en},
	urldate = {2025-03-08},
	publisher = {Springer},
	editor = {Lee, J. A. and Verleysen, M.},
	year = {2007},
}

@article{willems_paradigms_1991,
  author={Willems, J.C.},
  journal={IEEE Transactions on Automatic Control}, 
  title={Paradigms and puzzles in the theory of dynamical systems}, 
  year={1991},
  volume={36},
  number={3},
  pages={259-294},
}

@InProceedings{pmlr-v144-lian21a,
  title = {Nonlinear Data-Enabled Prediction and Control},
  author = {Lian, Y. and Jones, C. N.},
  booktitle = {Proceedings of the 3rd Conference on Learning for Dynamics and Control},
  pages = {523--534},
  year = {2021},
  volume = {144},
  series = {Proceedings of Machine Learning Research},
  month = {07 -- 08 June},
  publisher = {PMLR}
}

@article{huang_robust_2024,
	title = {Robust and {Kernelized} {Data}-{Enabled} {Predictive} {Control} for {Nonlinear} {Systems}},
	volume = {32},
	number = {2},
	journal = {IEEE Transactions on Control Systems Technology},
	author = {Huang, L. and Lygeros, J. and Dörfler, F.},
	month = mar,
	year = {2024},
	note = {Conference Name: IEEE Transactions on Control Systems Technology},
	pages = {611--624},
}

@misc{lazar_basis_2023,
	title = {Basis functions nonlinear data-enabled predictive control: {Consistent} and computationally efficient formulations},
	shorttitle = {Basis functions nonlinear data-enabled predictive control},
	publisher = {arXiv},
	author = {Lazar, M.},
	month = nov,
	year = {2023},
	note = {arXiv:2311.05360 [cs, eess, math]},
}

@article{elokda_data-enabled_2021,
	title = {Data-enabled predictive control for quadcopters},
	volume = {31},
	copyright = {© 2021 The Authors. International Journal of Robust and Nonlinear Control published by John Wiley \& Sons Ltd.},
	language = {en},
	number = {18},
	journal = {International Journal of Robust and Nonlinear Control},
	author = {Elokda, E. and Coulson, J. and Beuchat, P. N. and Lygeros, J. and Dörfler, F.},
	year = {2021},
	pages = {8916--8936},
}

@inproceedings{coulson_data-enabled_2019,
  author={Coulson, J. and Lygeros, J. and Dörfler, F.},
  booktitle={2019 18th European Control Conference (ECC)}, 
  title={Data-{Enabled} {Predictive} {Control}: {In} the {Shallows} of the {DeePC}}, 
  year={2019},
  volume={},
  number={},
  pages={307-312},
}

@article{berberich_linear_2022,
  author={Berberich, J. and Köhler, J. and Müller, M. A. and Allgöwer, F.},
  journal={IEEE Transactions on Automatic Control}, 
  title = {Linear tracking {MPC} for nonlinear systems {Part} {II}: {The} data-driven case},
  year={2022},
  volume={67},
  number={9},
  pages={4406-4421},
}

@article{dorfler_bridging_2023,
	title = {Bridging {Direct} and {Indirect} {Data}-{Driven} {Control} {Formulations} via {Regularizations} and {Relaxations}},
	volume = {68},
	number = {2},
	journal = {IEEE Transactions on Automatic Control},
	author = {Dörfler, F. and Coulson, J. and Markovsky, I.},
	month = feb,
	year = {2023},
	note = {Conference Name: IEEE Transactions on Automatic Control},
	pages = {883--897},
}

@incollection{goos_surprising_2001,
	address = {Berlin, Heidelberg},
	title = {On the {Surprising} {Behavior} of {Distance} {Metrics} in {High} {Dimensional} {Space}},
	volume = {1973},
	language = {en},
	booktitle = {Database {Theory} — {ICDT} 2001},
	publisher = {Springer Berlin Heidelberg},
	author = {Aggarwal, C. C. and Hinneburg, A. and Keim, D. A.},
	editor = {Goos, Gerhard and Hartmanis, Juris and Van Leeuwen, Jan and Van Den Bussche, Jan and Vianu, Victor},
	year = {2001},
	note = {Series Title: Lecture Notes in Computer Science},
	pages = {420--434},
}

@article{markovsky_data-driven_2023,
	title = {Data-{Driven} {Control} {Based} on the {Behavioral} {Approach}: {From} {Theory} to {Applications} in {Power} {Systems}},
	volume = {43},
	copyright = {https://ieeexplore.ieee.org/Xplorehelp/downloads/license-information/IEEE.html},
	shorttitle = {Data-{Driven} {Control} {Based} on the {Behavioral} {Approach}},
	language = {en},
	number = {5},
	journal = {IEEE Control Systems},
	author = {Markovsky, I. and Huang, L. and Dörfler, F.},
	month = oct,
	year = {2023},
	pages = {28--68},
}

@article{breschi_data-driven_2023,
	title = {Data-driven predictive control in a stochastic setting: a unified framework},
	volume = {152},
	shorttitle = {Data-driven predictive control in a stochastic setting},
	journal = {Automatica},
	author = {Breschi, V. and Chiuso, A. and Formentin, S.},
	month = jun,
	year = {2023},
	pages = {110961},
}

@article{tenenbaum_global_2000,
	title = {A {Global} {Geometric} {Framework} for {Nonlinear} {Dimensionality} {Reduction}},
	volume = {290},
	language = {en},
	number = {5500},
	journal = {Science},
	author = {Tenenbaum, J. B. and Silva, V. De and Langford, J. C.},
	month = dec,
	year = {2000},
	pages = {2319--2323},
}

@article{martinelli_data-driven_2022,
	title = {Data-{Driven} {Optimal} {Control} of {Affine} {Systems}: {A} {Linear} {Programming} {Perspective}},
	volume = {6},
	shorttitle = {Data-{Driven} {Optimal} {Control} of {Affine} {Systems}},
	journal = {IEEE Control Systems Letters},
	author = {Martinelli, A. and Gargiani, M. and Draskovic, M. and Lygeros, J.},
	year = {2022},
	note = {Conference Name: IEEE Control Systems Letters},
	pages = {3092--3097},
}

@article{willems_note_2005,
	title = {A note on persistency of excitation},
	volume = {54},
	number = {4},
	journal = {Systems \& Control Letters},
	author = {Willems, J. C. and Rapisarda, P. and Markovsky, I. and De Moor, B. L. M.},
	month = apr,
	year = {2005},
	pages = {325--329},
}

@article{berberich_overview_2024,
	title = {An {Overview} of {Systems}-{Theoretic} {Guarantees} in {Data}-{Driven} {Model} {Predictive} {Control}},
	language = {en},
	author = {Berberich, J. and Allgöwer, F.},
	month = oct,
	year = {2024},
	note = {Publisher: Annual Reviews},
}

@misc{kwiatkowski_gymnasium_2024,
	title = {Gymnasium: {A} {Standard} {Interface} for {Reinforcement} {Learning} {Environments}},
	shorttitle = {Gymnasium},
	publisher = {arXiv},
	author = {Kwiatkowski, A. and Towers, M. and Terry, J. and Balis, J. U. and Cola, G. De and Deleu, T. and Goulão, M. and Kallinteris, A. and Krimmel, M. and KG, A. and Perez-Vicente, R. and Pierré, A. and Schulhoff, S. and Tai, J. J. and Tan, H. and Younis, O. G.},
	month = oct,
	year = {2024},
	note = {arXiv:2407.17032},
}

@article{morari_model_1999,
	title = {Model predictive control: past, present and future},
	volume = {23},
	shorttitle = {Model predictive control},
	number = {4},
	journal = {Computers \& Chemical Engineering},
	author = {Morari, M. and Lee, J. H.},
	month = may,
	year = {1999},
	pages = {667--682},
}

@book{sutton_reinforcement_1998,
	address = {Cambridge, MA},
	title = {Reinforcement {Learning}: {An} {Introduction}},
	language = {en},
	publisher = {MIT Press},
	author = {Sutton, R. S and Barto, A. G},
	year = {1998},
}

@misc{cummins_deepc-hunt_2024,
	title = {{DeePC}-{Hunt}: {Data}-enabled {Predictive} {Control} {Hyperparameter} {Tuning} via {Differentiable} {Optimization}},
	shorttitle = {{DeePC}-{Hunt}},
	publisher = {arXiv},
	author = {Cummins, M. and Padoan, A. and Moffat, K. and Dorfler, F. and Lygeros, J.},
	month = dec,
	year = {2024},
	note = {arXiv:2412.06481 [math]},
}

@article{williams_model_2017,
	title = {Model {Predictive} {Path} {Integral} {Control}: {From} {Theory} to {Parallel} {Computation}},
	volume = {40},
	shorttitle = {Model {Predictive} {Path} {Integral} {Control}},
	language = {en},
	number = {2},
	journal = {Journal of Guidance, Control, and Dynamics},
	author = {Williams, G. and Aldrich, A. and Theodorou, E. A.},
	month = feb,
	year = {2017},
	pages = {344--357},
}

@inproceedings{nagabandi_neural_2018,
	title = {Neural {Network} {Dynamics} for {Model}-{Based} {Deep} {Reinforcement} {Learning} with {Model}-{Free} {Fine}-{Tuning}},
	booktitle = {2018 {IEEE} {International} {Conference} on {Robotics} and {Automation} ({ICRA})},
	author = {Nagabandi, A. and Kahn, G. and Fearing, R. S. and Levine, S.},
	month = may,
	year = {2018},
	pages = {7559--7566},
}

@misc{kurutach_model-ensemble_2018,
	title = {Model-{Ensemble} {Trust}-{Region} {Policy} {Optimization}},
	publisher = {arXiv},
	author = {Kurutach, T. and Clavera, I. and Duan, Y. and Tamar, A. and Abbeel, P.},
	month = oct,
	year = {2018},
	note = {arXiv:1802.10592 [cs]},
}

@misc{moerland_model-based_2022,
	title = {Model-based {Reinforcement} {Learning}: {A} {Survey}},
	shorttitle = {Model-based {Reinforcement} {Learning}},
	publisher = {arXiv},
	author = {Moerland, T. M. and Broekens, J. and Plaat, A. and Jonker, C. M.},
	month = mar,
	year = {2022},
	note = {arXiv:2006.16712 [cs]},
}

@incollection{botev_chapter_2013,
	series = {Handbook of {Statistics}},
	title = {Chapter 3 - {The} {Cross}-{Entropy} {Method} for {Optimization}},
	volume = {31},
	booktitle = {Handbook of {Statistics}},
	publisher = {Elsevier},
	author = {Botev, Z. I. and Kroese, D. P. and Rubinstein, R. Y. and L’Ecuyer, P.},
	editor = {Rao, C. R. and Govindaraju, Venu},
	month = jan,
	year = {2013},
	pages = {35--59},
}

@INPROCEEDINGS{berkenkamp_safe_2016,
  author={Berkenkamp, F. and Schoellig, A. P. and Krause, A.},
  booktitle={2016 IEEE International Conference on Robotics and Automation (ICRA)}, 
  title = {Safe {Controller} {Optimization} for {Quadrotors} with {Gaussian} {Processes}},
  year={2016},
  volume={},
  number={},
  pages={491-496},
}

@article{fefferman_testing_2016,
	title = {Testing the manifold hypothesis},
	volume = {29},
	language = {English},
	number = {4},
	journal = {Journal of the American Mathematical Society},
	author = {Fefferman, C. and Mitter, S. and Narayanan, H.},
	month = oct,
	year = {2016},
	pages = {983--1049},
}

\newpage
\appendix
\onecolumn
\section{DeePC}\label{app:dpc}
Data-enabled Predictive Control (DeePC) solves a receding horizon optimal control problem based purely on data for a linear time-invariant system (LTI) of the form
\begin{equation}
  y_\text{f}(k) = F_\text{p}\begin{bmatrix}u_{\text{p}}(k-1) \\ y_{\text{p}}(k-1) \end{bmatrix}  + F_\text{f}u_\text{f}(k),
  \label{eq:lti-system}
\end{equation}
where $u$ and $y$ are partitioned into ``past'' and ``future'' quantities $u_\text{p}, u_\text{f}$ and $y_\text{p}, y_\text{f}$ respectively. This means that the sequence of predicted future measurements, the past measurements used by the predictor, the sequence of past inputs applied to the system and the future control input sequence respectively given by
\begin{align*}
y_\text{f}(k) &\coloneqq \begin{bmatrix}y(k)^\top & \dots & y(k+T_\text{f}-1)^\top \end{bmatrix}^\top \in \mathbb{R}^{T_\text{f}\cdot p},\\
u_\text{f}(k) &\coloneqq \begin{bmatrix}u(k)^\top & \dots & u(k+T_\text{f}-1)^\top \end{bmatrix}^\top \in \mathbb{R}^{T_\text{f}\cdot m},\\
y_\text{p}(k-1) &\coloneqq \begin{bmatrix}y(k-T_\text{p})^\top & \dots & y(k-1)^\top \end{bmatrix}^\top \in \mathbb{R}^{T_\text{p}\cdot p} \text{, and}\\
u_\text{p}(k-1) &\coloneqq \begin{bmatrix}u(k-T_\text{p})^\top & \dots & u(k-1)^\top \end{bmatrix}^\top \in \mathbb{R}^{T_\text{p}\cdot m}.
\end{align*}
Let $\mathcal{U}$ and $\mathcal{Y}$ denote constraint sets on the input and output $u_\text{f}$ and $y_\text{f}$ respectively. Given a set of stage cost functions $c_i$ and past trajectory of the system, we are interested in resolving the following problem in a receding horizon fashion.  
\begin{subequations}\label{eq:mpc}
\begin{align}
    \min_{u_\text{f}, y_\text{f}} \quad& \sum_{i=0}^{T_\text{f}-1} c_i(u_{\text{f},i}, y_{\text{f},i})\\
    \text{s.t.} \quad & y_\text{f} = F_\text{p}\begin{bmatrix}u_{\text{p}}\\ y_{\text{p}} \end{bmatrix}  + F_\text{f}u_\text{f},\\
    & (u_\text{f}, y_\text{f}) \in \mathcal{{U} \times \mathcal{Y}}. 
\end{align}
\end{subequations}

Without access to the model \eqref{eq:lti-system}, we will need to reconstruct it from offline data. 
Let $v_T = \left\{ v(i)\right\}_{i=0}^{T-1}$ denote measurements of a signal $v$ with length
$T\in \mathbb{Z}_{\geq 1}$. We define the Hankel matrix of depth $L$ of $v_T$ as
\begin{equation}
  H_L(v_T) = 
  \begin{bmatrix}
    v(0) & v(1) & \dots & v(T - L - 1) \\
    v(1) & v(2) & \dots & v(T - L) \\
    \vdots & \vdots & \ddots & \vdots \\
    v(L-1) & v(L) & \dots & v(T-1)
  \end{bmatrix}
\end{equation}

Note that each column of the Hankel matrix contains a trajectory of length $L$.

\begin{definition}[Persistency of Excitation]
  For a given sequence $v_T$, we call the sequence \emph{persistently exciting of order $L$} if $H_L(v_T)$ has full row rank.
\end{definition}

Using persistently exciting inputs, Willems' Fundamental Lemma allows us to parametrize all finite-length trajectories of a controllable linear system. 
\begin{lemma}[The Fundamental Lemma \cite{willems_note_2005}]
  \label{lm:fundamental}
  Consider a controllable linear time-invariant system of order $n$. Given an input sequence $u_T$ which is
  persistently exciting of order $L\cdot m + n$ and the corresponding output sequence $y_T$, then there exists $g$ such
  that any admissible trajectory (u, y) of length $L$ can be expressed as a linear combination
  \begin{equation}
    \begin{bmatrix}
      H_{L}(u_T) \\
      H_{L}(y_T)
    \end{bmatrix}
    g
    = 
    \begin{bmatrix}
      u\\
      y
    \end{bmatrix}.
  \label{eq:hankel_matrix}
  \end{equation}
\end{lemma}

DeePC is based on the use of this implicit predictor \eqref{eq:hankel_matrix} as a proxy for the explicit model representation in \eqref{eq:lti-system}. In an online manner, DeePC matches the ``past'' data with the most recently seen inputs and outputs, and uses the ``future'' data to match a prediction over which we optimize. This results in a receding-horizon predictive controller based purely on data:
\begin{subequations}\label{eq:deepc}
\begin{align}
    \min_{u_\text{f}, y_\text{f},g} \quad& \sum_{i=0}^{T_\text{f}-1} c_i(u_{\text{f},i}, y_{\text{f},i}) + r(g)\\
    \text{s.t.} \quad & \begin{bmatrix} H_{T_\text{p}+T_\text{f}}(u_T) \\  H_{T_\text{p}+T_\text{f}}(y_T)\end{bmatrix}  g = \begin{bmatrix} u_{\text{p}}\\ u_{\text{f}} \\ y_{\text{p}}\\y_{\text{f}} \end{bmatrix},\label{eq:deepc-hankel}\\
    & (u_\text{f}, y_\text{f}) \in \mathcal{{U} \times \mathcal{Y}}.
\end{align}
\end{subequations}

Here, $r(g)$ denotes suitable regularizers on $g$ \cite{dorfler_bridging_2023} which improve robustness of the controller in the presence of noise. The implicit predictor in \eqref{eq:hankel_matrix} can readily be extended to affine systems as follows \cite{martinelli_data-driven_2022, berberich_linear_2022}
\begin{equation}\label{eq:affine_hankel}
  \begin{bmatrix} H_t(u_T) \\ H_t(y_T) \\ \mathbbm{1}^\top \end{bmatrix}g = \begin{bmatrix} u \\ y \\ 1\end{bmatrix}.
\end{equation}

\subsection{From Implicit to Explicit Predictor}
Partitioning the predictor \eqref{eq:affine_hankel} into past and future states results in
\begin{equation}\label{eq:partitioned_hankel}
  \begin{bmatrix}
    U_\text{p}\\
    U_\text{f}\\
    Y_\text{p}\\
    Y_\text{f}\\
    \mathbbm{1}^\top
  \end{bmatrix}
  g
  =
  \begin{bmatrix}
    u_\text{p}\\
    u_\text{f}\\
    y_\text{p}\\
    y_\text{f}\\
    1
  \end{bmatrix},
\end{equation}
which is an implicit predictor for the future measurement trajectory $y_\text{f}$ given past measurements $y_\text{p}$,
the corresponding past input sequence $u_\text{p}$ and a future input sequence $u_\text{f}$. Using the partitions
\begin{equation}
  H_z \coloneqq \begin{bmatrix}U_\text{p} \\ U_\text{f} \\ Y_\text{p} \\ \mathbbm{1}^\top \end{bmatrix},
  \ z \coloneqq \begin{bmatrix}
      u_\text{p} \\ u_\text{f} \\ y_\text{p} \\ 1
  \end{bmatrix},
\end{equation}
we can eliminate $g$ from the implicit predictor in \cref{eq:partitioned_hankel} and write the equivalent explicit predictor
\begin{equation}
  y_\text{f} = Y_\text{f} H_z^\dagger z + Y_\text{f}N_{H_z}g_0,
\end{equation}
where $A^\dagger$ denotes the pseudo-inverse of $A$, $N_{H_z}$ is the null-space projection matrix of $H_z$ and $g_0$ is an arbitrary perturbance to the least squares solution $g = H_z^\dagger z$.
Assuming $H_z$ has full row rank, then the
least squares solution (i.e. $N_{H_z}g_0 = 0$) is
\begin{equation}
  y_\text{f} = Y_\text{f}H_z^\top\left(H_z H_z^\top\right)^{-1}z.
  \label{eq:lls}
\end{equation}

\newpage
\section{SQP-MPC}\label{app:sqpmpc}
Nonlinear MPC tries to optimize a cost function subject to the state evolution of a nonlinear system of the form
\begin{equation}
  y_\text{f}(k) = f\left(u_{\text{p}}(k-1), u_\text{f}(k), y_{\text{p}}(k-1)\right).
  \label{eq:narx}
\end{equation}
If we had access to $f$, then we could use the receding-horizon optimal control problem
\begin{subequations}
\begin{align}\label{eq:nonlinocp}
    \min_{u_\text{f}, y_\text{f}} \quad& \sum_{i=0}^{T_\text{f}-1} c_i(u_{\text{f},i}, y_{\text{f},i})\\
    \text{s.t.} \quad & y_\text{f} = f\left(u_{\text{p}}, u_\text{f}, y_{\text{p}}\right),\label{eq:mpc-model}\\
    & (u_\text{f}, y_\text{f}) \in \mathcal{{U} \times \mathcal{Y}}
\end{align}
\end{subequations}
to obtain an optimal control input sequence $u^\star_\text{f}$. One method of solving a nonlinear optimal control problem of this form is using SQP, which repeatedly solves a linearization of the nonlinear problem \eqref{eq:nonlinocp}, resulting in subproblem \eqref{eq:iompc}. The solution of said subproblem is then used to update the estimate of the optimal open-loop solution $(\tilde{u}_\text{f},\tilde{y}_\text{f})$.
\begin{subequations}\label{eq:iompc}
\begin{align}
  \min_{\Delta u_\text{f},\Delta y_\text{f}} \quad & \sum_{i=0}^{T_\text{f}-1} 
      \begin{bmatrix}\Delta y_{\text{f},i}^\top &  \Delta u_{\text{f},i}^\top & 1\end{bmatrix} \tilde{H}_i
      \begin{bmatrix}\Delta y_{\text{f},i} \\  \Delta u_{\text{f},i}  \\ 1\end{bmatrix} \label{eq:iompc:cost}\\
  \text{s.t.} \quad & \Delta y_\text{f} = F_{\text{p}} \begin{bmatrix}\Delta u_\text{p}\\ \Delta y_\text{p}\end{bmatrix} + F_{\text{f}} \Delta u_\text{f}, \label{eq:iompc:linearization}\\
                    & \Delta z_\text{p} =\begin{bmatrix} u_\text{p}(k) - \tilde{u}_\text{p} \\ y_\text{p}(k) - \tilde{y}_\text{p} \end{bmatrix},\\
                    & (\Delta u_{\text{f},i}, \Delta y_{\text{f},i}) \in \mathcal{U}\times\mathcal{Y} \quad \forall i \in \left\{0, \dots, T_\text{f}-1\right\},
\end{align}
\end{subequations}
where the symbols $F_{\text{p}} \coloneqq \frac{\partial}{\partial \begin{bmatrix}u_\text{p}^\top & y_\text{p}^\top\end{bmatrix}^\top}f(u_\text{p}, y_\text{p}, \tilde{u}_\text{f})$ and
${F_{\text{f}} \coloneqq \frac{\partial}{\partial \tilde{u}_\text{f}}f(u_\text{p}, y_\text{p}, \tilde{u}_\text{f})}$
indicate the respective Jacobians of system~\eqref{eq:narx} evaluated at the current solution estimate
$(u_\text{p}, y_\text{p}, \tilde{u}_\text{f}, \tilde{y}_\text{f})$.

The full SQP algorithm can be summarized as follows \cite{diehl2002real}
\begin{algorithm}[H]
  \caption{SQP-MPC (IO Representation)}\label{alg:sqp-mpc}
\begin{algorithmic}[1]
  \FUNCTION{SQP-MPC($z_\text{p}$)}
    \WHILE{not converged}
      \STATE linearize dynamics, constraints and cost at $(z_\text{p}, \tilde{u}_\text{f}, \tilde{y}_\text{f})$
      \STATE Solve \cref{eq:iompc}
      \STATE $(\tilde{u}_\text{f}, \tilde{y}_\text{f}) \leftarrow (\tilde{u}_\text{f}, \tilde{y}_\text{f}) + (\Delta u_\text{f}, \Delta y_\text{f})$
    \ENDWHILE 
    \STATE Return $\tilde{u}_{\text{f},0}$
  \ENDFUNCTION
\end{algorithmic}
\end{algorithm}

\subsection{Equivalence between Select-DPC and SQP-MPC}
Select-DPC has a very natural interpretation as solving a
nonlinear MPC problem using SQP, where the analytical Jacobian computation is replaced by a data-driven approach
of estimating the Jacobian of the input-output behavior from data. Specifically, one can select data points ``close'' to the current linearization point (given by the open loop solution), resulting in an equivalent, data-driven approximation of $F_\text{p}$ and $F_\text{f}$ in \cref{eq:iompc} that approaches the analytical Jacobians assuming sufficient amounts of data in the neighborhood around the linearization point (i.e. $\operatorname{card}(\mathcal{D})$ tending to infinity).

\newpage
\section{Isomap}\label{app:isomap}
Isomap, short for Isometric Mapping, is a nonlinear dimensionality reduction technique introduced by Tenenbaum \cite{tenenbaum_global_2000}.
It can be seen as an extension of classical Multidimensional Scaling (MDS) where, instead of using Euclidean interpoint distances, geodesic distances between the data points are preserved. These geodesic distances are estimated using shortest distances on a connected graph which connects neighboring data points. These shortest paths are then arranged in a distance matrix $D$. Isomap then finds an embedding by applying classical MDS. Computing the distance of a new query data point to the data set in the embedding dimension is done by first linking the new data point into the neighborhood graph and then projecting the data point into the embedding space.

Isomap is able to capture the overall nonlinear nature of the data manifold and is guaranteed to produce a global optimizer of its reconstruction loss. Furthermore, due to preserving geodesic distances, points close in the input space are also close in the embedding space, resulting in interpretable results. This is especially relevant for Select-DPC which is build upon the idea of selecting data points according to their associated distance to a query point.

\begin{wrapfigure}[13]{r}{0.5\textwidth}
  \vspace{-0.5cm}
  \centering
    \includegraphics[scale=0.8]{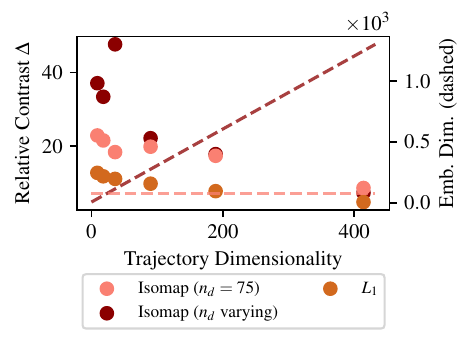}
  \caption{Relative contrast with Isomap embedding compared to $L_1$ selection.
  }
  \label{fig:rel_con_iso}
\end{wrapfigure}
The relative contrast of a data set ${\Delta = (d_\text{max} - d_\text{min})/d_\text{min}}$, with $d_\text{max/min}$ indicating the maximum and minimum norm values of the data set respectively, is a helpful surrogate to determine how distinct each data point appears under a given distance metric.
\Cref{fig:rel_con_iso} shows the relative contrast of a data set collected from the rocket simulator in \cref{sec:results} as a function of trajectory dimensionality when
embedded using Isomap. It compares relative contrast given a fixed embedding dimensionality as well as a varying one,
which increases according to $n + T_\text{f}\cdot m$, to the relative contrast of the $L_1$ norm. Both Isomap methods have better relative contrast compared to $L_1$ selection. While a
varying embedding dimensionality offers greater relative contrast for small trajectory lengths, we encounter
diminishing returns for large trajectory lengths.

\subsection{Isomap Parameter Selection}
\begin{figure}[h]
  \begin{center}
    \includegraphics[scale=1]{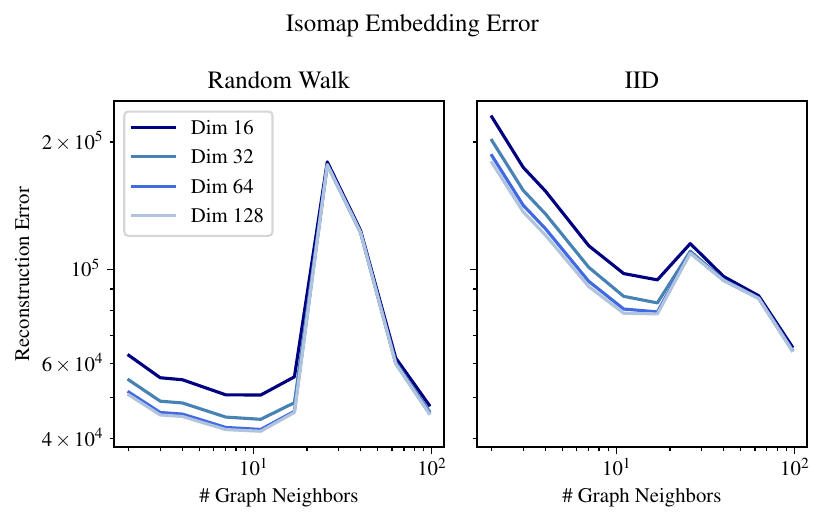}
  \end{center}
  \caption{Isomap reconstruction error as a function a embedding dimensionality and number of neighbors used in the
  adjacency graph.}
  \label{fig:reconstruction_error}
\end{figure}

Isomap comes with a set of hyper parameters that need to be selected for the data set at hand. Specifically, the number of closest data points that should be considered as neighbors during the neighborhood graph construction and the dimensionality of the embedding space.
\Cref{fig:reconstruction_error} shows the reconstruction error of the data sets as a function of number of graph neighbors used during adjacency graph construction and number of dimensions of the embedding space.
The graphs show a minimum around 10 graph neighbors while we can observe diminishing returns from increasing the embedding dimensionality
from 64 to 128 which is consistent with the formula $T_\text{f}\cdot m + n$ which suggests that the locally linear system has a dimensionality of 96.

Furthermore, note the sharp increase in reconstruction error as the number of graph neighbors increases after the local minimum. This demonstrates the effect of bad hyperparameter selection
which can lead to short-circuiting or over-connectedness in the graph, leading to worse embedding performance as the local manifold structure is lost \cite{tenenbaum_global_2000}.

\newpage
\section{Experiment Hyperparameters}\label{app:hyperparams}
All experiments were performed using a quadratic tracking cost of the form
\begin{equation}
    c_i(u_{\text{f},i}, y_{\text{f},i}) = \|y_{\text{f},i} - y_r\|_Q + \|u_{\text{f},i}\|_R,
\end{equation}
where $Q$ and $R$ are positive (semi-)definite cost matrices and $y_r$ is a reference setpoint. Furthermore, for the regularization term $r(g)$ we use 1-norm and projection regularization \cite{markovsky_data-driven_2023} with weights $\lambda_1$ and $\lambda_\Pi$ respectively. The following tables shows the hyperparameters for the three sets of simulations.

\begin{table}[h]
    \centering
    \begin{tabular}{ c| c }
        Parameter & Value\\
        \hline
        convergence criterion $n_\text{max}$ & 1 \\ 
        $Q$  &  $\operatorname{diag}\left(\begin{bmatrix} 40 & 20 & 20 & 1 & 3000 & 30\end{bmatrix}\right)$\\
        $R$ & $\operatorname{diag}\left(\begin{bmatrix} 10 & 10 & 10\end{bmatrix}\right)$\\
        $\lambda_1$   &  0 \\
        $\lambda_\pi$ &  5000
    \end{tabular}
    \caption{Hyperparameters for the rocket simulation environment.}
    \label{tab:hyperparams-rocket}
\end{table}
\begin{table}[h]
    \centering
    \begin{tabular}{ c| c }
        Parameter & Value\\
        \hline
        convergence criterion $n_\text{max}$ & 5\\
        $Q$ &$\operatorname{diag}\left(\begin{bmatrix} 0 & 0 & 0 & 0 & \num{40000} & \num{40000} & 10 & 10\end{bmatrix}\right)$\\
        $R$ & $\operatorname{diag}\left(\begin{bmatrix} 10 & 10\end{bmatrix}\right)$\\
        $\lambda_1$ & 10\\
        $\lambda_\pi$ & \num{10000
        }
    \end{tabular}
    \caption{Hyperparameters for the reacher simulation environment.}
    \label{tab:hyperparams-reacher}
\end{table}
\begin{table}[H]
    \centering
    \begin{tabular}{ c| c }
    Parameter & Value\\
    \hline
        convergence criterion $n_\text{max}$ & 1 \\
        $Q$ & $\operatorname{diag}\left(\begin{bmatrix} 5 & 0 & \num{10000} & 0.1 & 0.1\end{bmatrix}\right)$\\
        $R$ & $1$\\
        $\lambda_1$   &  \num{50000}\\
        $\lambda_\pi$ & 0
    \end{tabular}
    \caption{Hyperparameters for the cart-pole simulation environment.}
    \label{tab:hyperparams-pendulu,}
\end{table}

\end{document}